\documentstyle[11pt]{article}
\newlength{\bredde}
\def\slash#1{\settowidth{\bredde}{$#1$}\ifmmode\,\raisebox{.15ex}{/}
\hspace*{-\bredde} #1\else$\,\raisebox{.15ex}{/}\hspace*{-\bredde} #1$\fi}
\newcommand{\beq}{\begin{equation}}
\newcommand{\eeq}{\end{equation}}
\newcommand{\bea}{\begin{eqnarray}}
\newcommand{\eea}{\end{eqnarray}}
\newcommand{\noi}{\vspace{12pt}\noindent}
\newcommand{\lG}{\raise.3ex\hbox{$\stackrel{\leftarrow}{G}$}}
\newcommand{\lU}{\raise.3ex\hbox{$\stackrel{\leftarrow}{U}$}}
\newcommand{\lP}{\raise.3ex\hbox{$\stackrel{\leftarrow}{P}$}}
\newcommand{\lp}{\raise.3ex\hbox{$\stackrel{\leftarrow}{P}$}}
\newcommand{\lPP}{\raise.3ex\hbox{$\stackrel{\leftarrow}{\cal P}$}}
\newcommand{\leta}{\raise.3ex\hbox{$\stackrel{\leftarrow}{\eta}$}}
\newcommand{\lOmega}{\raise.3ex\hbox{$\stackrel{\leftarrow}{\Omega}$}}
\newcommand{\ldr}{\raise.3ex\hbox{$\stackrel{\leftarrow}{\delta^r}$}}
\def\beqn{\begin{eqnarray}}
\def\eeqn{\end{eqnarray}}
\def\sepand{\rule{14cm}{0pt}\and}
\def\gtwid{\raise.3ex\hbox{$>$\kern-.75em\lower1ex\hbox{$\sim$}}}
\def\ltwid{\raise.3ex\hbox{$<$\kern-.75em\lower1ex\hbox{$\sim$}}}

\begin{document}
%\begin{Ntitlepage}
\topmargin -1.4cm
\oddsidemargin -0.8cm
\evensidemargin -0.8cm
\title{\Large{BRST Formulation of Partition Function Constraints}}

\vspace{0.5cm}

\author{{\sc Jorge Alfaro} \\
Fac. de Fisica \\ Universidad Catolica de Chile\\
Casilla 306, Santiago 22 \\Chile \\
\sepand
{\sc Klaus Bering}\\
Institute of Theoretical Physics\\Uppsala University\\P.O. Box 803\\
S-751 08 Uppsala\\Sweden\\
\sepand
{\sc Poul H. Damgaard}\\
The Niels Bohr Institute\\ Blegdamsvej 17\\ DK-2100 Copenhagen\\
Denmark} 
\maketitle
\vfill
\begin{abstract} We show that constraints on the generating functional
have direct BRST-extensions in terms of nilpotent operators $\Delta$
that annihilate this generating functional, and which may be of
arbitrarily high order. The free energy $F$ in the
presence of external sources thus satisfies a ``Master Equation'' which
is described in terms of a tower of higher antibrackets.
\end{abstract}
\vfill
%\vspace{6.2cm}
\begin{flushleft}
UUITP-23/96\\
NBI-HE-96-52 \\
hep-th/9610083
\end{flushleft}
\newpage
%\phantom{}
%\vfill
%\eject

%\setcounter{page}{1}

%\section{Introduction}
%\noindent
\noi
Recent work on the implementation of Ward Identities in the Lagrangian
path integral has shown that a novel antibracket structure naturally
emerges \cite{AD1,AD2,BDA}. Central in this construction is a non-Abelian 
``classical'' BRST operator $\Delta$, which in a precise sense provides a
non-Abelian generalization of the conventional $\Delta$-operator of
the Batalin-Vilkovisky formalism \cite{BV}. 

\noi
With the usual assignments of fields and antifields, the $\Delta$-operator
of the Batalin-Vilkovisky formalism is a 2nd order differential operator.
It can be understood as the Hamiltonian quantum BRST operator of Abelian
shift transformations in the ghost momentum representation \cite{AD2}.
Choosing instead $\Delta$ as the Hamiltonian quantum BRST charge of more
general (non-Abelian) transformations, one obtains, in the same 
ghost momentum representation, the corresponding non-Abelian generalization.
It is a differential operator in fields and antifields of arbitrarily 
high order. Because the BRST operator is nilpotent by construction, so is the 
generalized $\Delta$-operator. Just as the usual 2nd order $\Delta$-operator 
of the Batalin-Vilkovisky formalism can be used to impose correct
Schwinger-Dyson equations in the path integral, the classical
non-Abelian $\Delta$-operators can be used to impose different Ward 
Identities \cite{AD1}. Higher-order $\Delta$-operators were actually first
considered, by Batalin and Tyutin \cite{BT}, in the context of quantum
deformations of the usual 2nd-order formalism. Very recently, a Lagrangian
path integral formulation has also been given \cite{BBD}.

\noi
Because the non-Abelian $\Delta$-operators can be differential operators
of higher order than 2, they generate not just the usual antibracket
$(A,B)$ (viewed in this context, up to a sign, as a two-bracket $\Phi^2(A,B)$),
but a whole hierarchy of higher antibrackets $\Phi^n(A_1,\ldots,A_n)$.
The mathematical structure behind this sequence of higher antibrackets
was developed some time ago by Koszul \cite{Koszul}, and has more 
recently been considered by Akman \cite{Akman}.

\noi
The purpose of this short note is to show that also the Hamiltonian 
quantum BRST operator $\Omega$ in the ghost {\em coordinate} 
representation can be given an interesting interpretation from the 
point of view of the Lagrangian path integral. 
Whereas the ghost momentum representation provides us with a
$\Delta$-operator which defines the quantum Master Equation for the action  
$S$, we shall show that the ghost coordinate representation of $\Omega$ 
gives the appropriate operator for defining 
the analogous quantum Master Equation
for the {\em free energy} $F$ in the presence of external sources.
Phrased otherwise, we will see how constraints imposed by the path
integral itself can be formulated in terms of BRST constraints $\Omega$ on the
partition function \mbox{${\cal Z}={\cal Z}[j,\eta]$} 
(in the presence of suitable sources $j$ and $\eta$). 
\beq
{\cal Z}[j,\eta]~ \lOmega ~=~ 0 ~.
\eeq
A first example of this was the BRST formulation 
of Virasoro constraints in matrix models \cite{AJ}, 
and a brief account of its generalization to more general
settings has been presented by one of the authors in ref. \cite{Alfaro}.
The formulation of low-dimensional string theory in terms of a
Batalin-Vilkovisky--like quantum Master Equation acting on the free
energy $F$ \cite{Verlinde} is related to this idea as well.

\noi
Our treatment will be based on a combination of both BRST and anti-BRST
symmetries. We do not believe this is essential, but it certainly
simplifies the derivation. Although the final
results will not be $Sp(2)$-covariant, we also prefer to combine the
symmetries into an extended $Sp(2)$ symmetry. There will
thus be two associated variations $\delta^a$, with $a =
1, 2$. Indices will be raised and lowered with the help of 
$\varepsilon_{ab}$
(and its inverse), the invariant tensor in $Sp(2)$. We require that
\beq
\delta^a\delta^b + \delta^b\delta^a ~=~ 0 ~,
\eeq
the $Sp(2)$-generalization of nilpotency. It is also convenient to 
introduce the commutator of the two symmetries,
\beq
d ~\equiv~ {\scriptstyle \frac{1}{2}} \varepsilon_{ab}\delta^a\delta^b ~.
\eeq

\noi
We assume that the fields $\phi^A$ with Grassmann grading 
\mbox{$\epsilon_A \equiv\epsilon(\phi^A)$} 
transform under a local non-Abelian symmetry 
group $G$:\footnote{Space-time integrations are always
implicitly implied in the sums over repeated indices.}
\beq
      \phi^A\longrightarrow  g^A{}_B \ \phi^B
~=~\phi^B \ g_B{}^A  ~.\label{contratransf}
\eeq
Here we have introduced the super transposed matrix 
\beq
   g_B{}^A=(-1)^{(\epsilon_A+1)\epsilon_B} g^A{}_B~.
\eeq
The matrix $g^A{}_B$ is assumed not to depend on the field $\phi^A$.
A simple (bosonic) example is a $U(N)$ matrix model under the 
(global) adjoint action of $G=U(N)$.
Then the index $A$ is a double index: \mbox{$\phi^A=\phi^{a_1}{}_{a_2}$}, 
and if $g^{a}{}_{b}$ is given in the fundamental representation of  $U(N)$, 
then 
\beq
g^A{}_B=g^{a_1}{}_{b_1}  \ (g^{-1})^{b_2}{}_{a_2}  \  
=g^{a_1}{}_{b_1} \  (g^{\dagger})^{b_2}{}_{a_2} \ ~.
\eeq
Let us parametrize the action $S[\phi]$ in terms of a perhaps 
infinite set of group invariants $\chi_{\alpha}(\psi)$ and appropriate weights 
(sources) $j^{\alpha}$. That is,
\beq
S[\phi] ~=~ \chi_{\alpha}(\phi)j^{\alpha} ~.
\label{origaction}
\eeq 
The $\chi_{\alpha}$'s will be invariant under the local 
transformations generated by the non-Abelian group $G$. 
It is implicitly assumed here that $\{\chi_{\alpha}\}$ 
forms a complete (perhaps overcomplete) set of $G$-invariants.  

\noi
We take the path integral measure for the fields $\phi^A$ to be invariant 
under arbitrary local shifts \mbox{$\phi^A \to \phi^A+ \epsilon^A$} as 
well as under the
$G$-transformations (\ref{contratransf}). Consider now the following
(right) BRST transformations:\footnote{We omit an infinitesimal 
Grassmann-odd parameter, which we otherwise would have put 
to the right. This defines our rules of signs relating actual variations 
with these BRST transformations.}
\bea
\delta^a\phi^A &~=~& \psi^{Aa} \cr
\delta^a\psi^{Ab} &~=~& 0 ~.\label{BRST}
\eea
$\psi^{Aa}$ are ghosts and anti-ghosts. They will transform under the group 
$G$ as the original fields $\phi^A$, cf.\ (\ref{contratransf}), so that 
the $G$-transformations and BRST-transformations commute.
If one chooses a formulation without auxiliary fields, these are the
$Sp(2)$-symmetric generalizations \cite{sp2} of Schwinger-Dyson BRST
transformations \cite{AD0}. The Ward Identities of the above symmetry are 
of course only Schwinger-Dyson equations for theories with action $S = 0$.
These transformations simply express the translational invariance of the 
path integral measure.  We shall also introduce a spacetime-dependent 
matrix \mbox{$M_{AB}$} of Grassmann parity
\beq
\epsilon(M_{AB}) = \epsilon_A + \epsilon_B ~.
\eeq
The matrix $M$ is chosen to be graded-symmetric,
\beq
M_{AB} ~=~ (-1)^{\epsilon_A\epsilon_B+\epsilon_A+\epsilon_B}M_{BA} ~,
\eeq
and invertible. We denote its inverse by raised indices: $M^{AB}$.
$M_{AB}$ transforms as a covariant 2-tensor under the $G$-transformations, 
\beq
      M_{AB}  \longrightarrow   
 (g^{-1})_A{}^B \ M_{BC}  \  (g^{-1})^C{}_D ~.\label{2cotransf}
\eeq 
while $M^{AB}$ transforms as a contravariant 2-tensor.
The matrix $M$ is assumed not to depend on the field $\phi^A$.

\noi
Consider now the path integral weighted in addition by the $Sp(2)$-invariant
and $G$-invariant ghost-antighost action
\beq
S_0[\psi]  ~\equiv~ {\scriptstyle \frac{1}{2}}d(\phi^A M_{AB}\phi^B) 
 ~=~  {\scriptstyle \frac{1}{2}}(-1)^{\epsilon_A+1}
\varepsilon_{ab}\psi^{Aa}M_{AB}\psi^{Bb} ~.
\eeq
The functional integral can now be integrated over the ghost-antighost fields
with a well-defined Boltzmann factor. (Apart from this fact, the particular
measure $[d\psi]$ is not of interest at this point). We also add 
sources $\eta^{\alpha}_a$ for the $Sp(2)$ transformations of all the 
$G$-invariants. The result is the generating functional
\beq
{\cal Z}[j,\eta] 
~=~ \int[d\psi][d\phi]\exp\left[{\scriptstyle\frac{i}{\hbar}}\left(S_0[\psi]
+ \chi_{\alpha}(\phi)j^{\alpha} + \eta^{\alpha}_a\delta^a\chi_{\alpha}  
\right)\right] ~.
\eeq
Of course, to regain the original path integral with action (\ref{origaction}), 
we must set sources $\eta^{\alpha}_a$ equal to zero, 
and integrate out the ghost-antighost
fields $\psi^{Aa}$. This yields an irrelevant well-defined overall factor,
whose inverse we could choose to incorporate in the ghost-antighost
measure.

\noi
Let us next express the BRST Ward Identities of the above path integral
in a form that exploits the particular form of the action and sources. 
We use the fact that both $S_0$ and the functional measure are invariant
under the trivial BRST transformations (\ref{BRST}). Consider
\beq
0  ~=~ {\scriptstyle \frac{\hbar}{i}} \langle \delta_a\left(e^{\frac{i}{\hbar}
\left(\chi_{\alpha}j^{\alpha} + \eta^{\alpha}_b\delta^b
\chi_{\alpha}\right)}\right)\rangle_{0} 
 ~=~  \langle (-1)^{\epsilon_{\alpha}}j^{\alpha}(\delta^b\chi_{\alpha}) +
\eta^{\alpha}_c(\delta^b\delta^c\chi_{\alpha})\rangle\varepsilon_{ba} ~,
\label{Sp2WI}
\eeq
where $\epsilon_{\alpha}$ denotes the Grassmann parity of $\chi_{\alpha}$.
The subscript "$0$" in $\langle \cdots\rangle_{0}$ denotes 
that the Boltzmann factor  is \mbox{$e^{\frac{i}{\hbar}S_0}$}.
To keep the following derivation as simple as possible we now manifestly 
break $Sp(2)$ symmetry by noting that the term with $c \neq a$ does not 
contribute to (\ref{Sp2WI}):
\beq
 \langle (-1)^{\epsilon_{\alpha}}j^{\alpha}(\delta^b\chi_{\alpha}) +
\eta^{\alpha}_a(\delta^b\delta^a\chi_{\alpha})\rangle\varepsilon_{ba}~=~0 ~.
\label{WI}
\eeq
(No sum over $a$.)
This identity implicitly expresses a constraint on the path integral due to 
the symmetry of the measure under local shifts, and in this sense is analogous
to a general Schwinger-Dyson equation. The constraint can be made more
explicit by making use of the particular couplings to the source terms.
We have
\bea
\delta^a\chi_{\alpha} &~=~& \frac{\delta^r\chi_{\alpha}}{\delta\phi^A}
\psi^{Aa} ~=~ (-1)^{\epsilon_A+\epsilon_{\alpha}}\psi^{Aa}\frac{\delta^l
\chi_{\alpha}}{\delta\phi^A} \cr
\delta^b(\delta^a\chi_{\alpha}) &~=~& (-1)^{\epsilon_A+1}\frac{\delta^r
\delta^r \chi_{\alpha}}{\delta\phi^B\delta\phi^A}\psi^{Bb}\psi^{Aa} ~.
\label{transfchi}
\eea
The idea is now to replace the ghost-antighost pieces in the identity
(\ref{WI}) by terms acting on the external sources. Note that
\bea
\frac{\delta^r S_0}{\delta\psi^{Aa}} &=& (-1)^{\epsilon_B+1}\psi^{Bb}
M_{BA}\varepsilon_{ba} \cr
\frac{\delta^r S_{\eta}}{\delta\psi^{Aa}} &=& \frac{\delta^l \chi_{\alpha}}
{\delta\phi^A}\eta^{\alpha}_a ~,
\eea
with $S_{\eta} \equiv \eta^{\alpha}_a\delta^a\chi_{\alpha}$. Therefore,  
inside the path integral (\ref{WI}) we can by partial integration 
replace a single factor of  $\psi^{Aa}$ as follows:
\bea
\psi^{Aa} ~=~ \varepsilon^{ab}M^{AB}\frac{\delta^r S_0}{\delta\psi^{Bb}}
&~\to~& -\varepsilon^{ab}M^{AB} 
\left(\frac{\delta^r S_{\eta}}{\delta\psi^{Bb}} 
+ \frac{\hbar}{i}  \frac{\delta^r}{\delta\psi^{Bb}}  \right) \cr
&~=~& -\varepsilon^{ab}M^{AB} 
\left(\frac{\delta^l\chi_{\alpha}}{\delta\phi^A}\eta^{\alpha}_a 
+ \frac{\hbar}{i}  \frac{\delta^r}{\delta\psi^{Bb}}  \right)~,
\eea 
where the quantum term 
\mbox{$\frac{\hbar}{i}  \frac{\delta^r}{\delta\psi^{Bb}}$} is supposed to 
act on every ghost $\psi^{Aa}$ not in the Boltzmann factor.
If we therefore define a ``metric'' $d_{\alpha\beta}$ by
\beq
d_{\alpha\beta} ~\equiv~ \left(\frac{\delta^r\chi_{\alpha}}{\delta\phi^A}
\right)M^{AB}\left(\frac{\delta^l\chi_{\beta}}{\delta\phi^B}\right) ~,
\eeq
then the first term in eq. (\ref{WI}) reads
\beq
(-1)^{\epsilon_{\alpha}}j^{\alpha}\delta^b\chi_{\alpha}\varepsilon_{ba} 
~\to~ \eta^{\alpha}_a d_{\alpha\beta} j^{\beta}~.
\eeq

\noi
The second term in (\ref{WI}) contains two factors of ghosts $\psi$ (cf.\ 
(\ref{transfchi})). 
Substituting the first $\psi$ factor yields:
\beq
\eta^{\alpha}_a(\delta^b\delta^a\chi_{\alpha}) \varepsilon_{ba}
~\to~ (-1)^{A+\alpha} \psi^{Aa} 
\frac{\delta^r\delta^l\chi_{\alpha}}{\delta\phi^B\delta\phi^A}M^{BC}
\frac{\delta^l\chi_{\beta}}{\delta\phi^C}
\eta^{\beta}_a \eta^{\alpha}_a  -  i\hbar e_{\alpha}\eta^{\alpha}_a~,
\label{secondterm}
\eeq
where we have defined 
\beq
e_{\alpha} ~\equiv~ 
\frac{\delta^r\delta^r \chi_{\alpha}}{\delta\phi^A\delta\phi^B} M^{AB} 
~=~M^{AB} \frac{\delta^l\delta^l \chi_{\alpha}}{\delta\phi^B
\delta\phi^A} ~.
\eeq
To eliminate the second $\psi$ factor, we need to be more specific
regarding the types of transformations.

\noi
Up to this point we have made no assumptions about the $\chi_{\alpha}$'s, 
and in fact not exploited the idea of group invariants. Since
$\frac{\delta}{\delta\phi^A}$ is a covariant vector with respect to
the group $G$,
\beq
      \frac{\delta}{\delta\phi^A} \longrightarrow   
 \frac{\delta}{\delta\phi^A}  \ (g^{-1})^A{}_B ~,\label{cotransf}
\eeq
it follows that $d_{\alpha\beta}$ and $e_{\alpha}$ are invariants
themselves, $i.e.$ expressible in terms of $\chi_{\alpha}$'s: 
\mbox{$d_{\alpha\beta}= d_{\alpha\beta}(\chi)$} 
and \mbox{$e_{\alpha} = e_{\alpha}(\chi)$}.  
Furthermore, let us assume that there exists "structure constants" 
$f^{\alpha}_{\beta\gamma}$ so that 
\beq
\frac{\delta^r\delta^l\chi_{[\alpha}}{\delta\phi^B\delta\phi^A}M^{BC}
\frac{\delta^l\chi_{\beta]}}{\delta\phi^C}
~\equiv~ 
\frac{\delta^r\delta^l\chi_{\alpha}}{\delta\phi^B\delta\phi^A}M^{BC}
\frac{\delta^l\chi_{\beta}}{\delta\phi^C}
 - (-1)^{\epsilon_{\alpha}\epsilon_{\beta}}
\frac{\delta^r\delta^l\chi_{\beta}}
{\delta\phi^B\delta\phi^A}M^{BC}
\frac{\delta^l\chi_{\alpha}}{\delta\phi^C} 
~=~ -\frac{\delta^l\chi_{\gamma}} {\delta\phi^A} f^{\gamma}_{\alpha\beta}~.
\label{algebraeq}
\eeq
The structure constants
$f^{\alpha}_{\beta\gamma} = f^{\alpha}_{\beta\gamma}(\chi)$ are 
functions of the invariants only, as indicated. Their symmetry 
property is as follows: 
\beq
f^{\alpha}_{\beta\gamma} = 
- (-1)^{\epsilon_{\beta} \epsilon_{\gamma}} f^{\alpha}_{\gamma\beta}~.
\eeq
They satisfy a generalized Jacobi identity \cite{Batalin}:
\beq
\sum_{{\rm cycl.} ~\alpha, \beta,\gamma} 
(-1)^{\epsilon_{\alpha} \epsilon_{\gamma}}
\left( \frac{\delta^r f^{\epsilon}_{\alpha\beta}}{\delta \chi_{\delta}} 
d_{\delta\gamma}  
+ f^{\epsilon}_{\alpha\delta} f^{\delta}_{\beta\gamma}\right)~=~0~.
\eeq
For completeness we mention a simple consequence of 
\mbox{$d_{\alpha\beta}= d_{\alpha\beta}(\chi)$} being a group invariant:
\beq
\frac{\delta^r\delta^l\chi_{\{\alpha}}{\delta\phi^B\delta\phi^A}M^{BC}
\frac{\delta^l\chi_{\beta\}}}{\delta\phi^C}
~\equiv~ 
\frac{\delta^r\delta^l\chi_{\alpha}}{\delta\phi^B\delta\phi^A}M^{BC}
\frac{\delta^l\chi_{\beta}}{\delta\phi^C}
 + (-1)^{\epsilon_{\alpha}\epsilon_{\beta}}
\frac{\delta^r\delta^l\chi_{\beta}}
{\delta\phi^B\delta\phi^A}M^{BC}
\frac{\delta^l\chi_{\alpha}}{\delta\phi^C} 
~=~ \frac{\delta^l\chi_{\gamma}}{\delta\phi^A}
 \frac{\delta^l d_{\alpha\beta}}{\delta\chi_{\gamma}}~.
\eeq
This means that
\beq
d^{\gamma}_{\alpha\beta} ~\equiv~ 
\frac{\delta^l d_{\alpha\beta}}{\delta\chi_\gamma}
\eeq
are the ``symmetric'' structure constants. They too can be functions of
the invariants $\chi_{\alpha}$. The condition (\ref{algebraeq}) is
far from trivial, and at the heart of the present construction. It requires
that the $\delta^l\chi_{\alpha}/\delta\phi^A$ form a complete set of
covariant vectors. This may fail, for example if the fields do not transform
under an irreducible representation of the invariance group $G$.

\noi
Let us now proceed with the elimination of $\psi$ in the second term 
(\ref{secondterm}). {}From (\ref{algebraeq}), the classical piece becomes
\bea
 (-1)^{A+\alpha} \psi^{Aa} 
\frac{\delta^r\delta^l\chi_{\alpha}}{\delta\phi^B\delta\phi^A}M^{BC}
\frac{\delta^l\chi_{\beta}}{\delta\phi^C}
\eta^{\beta}_a \eta^{\alpha}_a 
~&=&~
 - {\scriptstyle \frac{1}{2}}  (-1)^{A+\alpha} 
\psi^{Aa} \frac{\delta^l\chi_{\gamma}} {\delta\phi^A} 
f^{\gamma}_{\alpha\beta}\eta^{\beta}_a \eta^{\alpha}_a  \cr
~&=&~
 - {\scriptstyle \frac{1}{2}}  (-1)^{\alpha} \left( \delta^a \chi_{\gamma} 
\right) 
f^{\gamma}_{\alpha\beta}\eta^{\beta}_a \eta^{\alpha}_a \cr
~&=&~
+ {\scriptstyle \frac{1}{2}}  (-1)^{\alpha} 
\frac{\delta^r S_{\eta}}{\delta \eta^{\gamma}_a} 
f^{\gamma}_{\alpha\beta}\eta^{\beta}_a \eta^{\alpha}_a ~.
\eea

\noi
Next, introduce the non-Abelian constraints, acting to the left,
\beq
\lG_{\alpha} ~\equiv~ d_{\alpha\beta}(\lp) j^{\beta} - i\hbar e_{\alpha}(\lp) ~,
\eeq
with the momentum operator conjugate to the sources acting to the left,
\bea
\lp_{\alpha} ~&\equiv&~ -i\hbar \frac{\ldr}{\delta j^{\alpha}} \cr
\lPP{}^a_{\alpha} ~&\equiv&~ -i\hbar \frac{\ldr}{\delta \eta^{\alpha}_a}~.
\eea 
One verifies that these constraints satisfy the algebra
\beq
[\lG_{\alpha},\lG_{\beta}] ~=~ -i\hbar \lG_{\gamma}f^{\gamma}_{\alpha\beta} ~.
\eeq
Note that the right hand side is proportional to $\hbar$, so that classically
this algebra is Abelian.

\noi
Associated with the constraints $\lG_{\alpha}$ is the quantum mechanical
BRST operator (also acting to the left),
\beq
\lOmega_a \equiv \lG_{\alpha}\eta^{\alpha}_a 
+{\scriptstyle \frac{1}{2}} (-1)^{\alpha}\lPP{}^a_{\gamma}
 f^{\gamma}_{\alpha\beta}(\lp)\eta^{\beta}_a
\eta^{\alpha}_a ~.
\eeq
This operator is nilpotent, due to the generalized Jacobi identity:
\beq
[\lOmega_a,\lOmega_a] = 0
\eeq 
(No sum over $a$.) In its present form the construction fails to be 
$Sp(2)$ symmetric, but is merely symmetric under two separate 
BRST symmetries, whose mutual relation is complicated. 
The identity (\ref{WI}) can now be expressed directly as a constraint
on the partition function:
\beq
{\cal Z}[j,\eta]~ \lOmega_a ~=~ 0 ~.
\eeq
Alternatively, if we define the free energy $F$ in the presence of sources by
$\exp(\frac{i}{\hbar}F[j,\eta])\equiv {\cal Z}$, then the free energy satisfies 
a
quantum Master Equation of the form
\beq
\Delta_a \exp \left(\frac{i}{\hbar} F \right) ~=~ 0 ~,\label{ME}
\eeq 
where we have defined a $\Delta$-operator acting to the right by
 $\Delta_a A \equiv A \lOmega_a$.\footnote{Since 
we are only dealing with one of the symmetries
corresponding to $a = 1, 2$, the subscript $a$ can be suppressed here.
We have not considered the possibility of imposing the
constraints in an $Sp(2)$-symmetric manner.}

\noi
Depending on the order $N$ of $\Delta_a$, the quantum Master Equation can
be written as a sum over antibrackets $\Phi^n_{\Delta_{a}}$ of order less
than and equal to $N$ \cite{AD2,BDA}:
\beq
\sum_{k=1}^N \left(\frac{i}{\hbar}\right)^k\frac{1}{k!}\Phi^k_{\Delta_{a}}
(F,\ldots,F) ~=~ 0 ~.
\eeq
A particular example is provided by the Virasoro constraints, as already
discussed in ref. \cite{AJ}. There the metric $d_{\alpha\beta}(P)$ is linear
in $P$, and $e_{\alpha}(P)$ is quadratic. The BRST operator generating
Virasoro constraints thus happens to be of second order. As should be obvious
from the above, one can however easily find examples where the relevant
operator is of higher order.
 
\vspace{0.5cm}

\noindent
{\sc Acknowledgement:}~
The work of J.A. is partially supported by Fondecyt 1950809 and a 
collaboration CONACYT(M\'{e}xico)-CONICYT. The work of K.B. and P.H.D.
has been partially supported by NorFA grants No. 95.30.182-O 
and 96.15.053-O, respectively.

\vspace{0.5cm}
%\newpage

\end{document}